\documentclass[11pt]{article}
\usepackage[margin=1in]{geometry}
\usepackage[table]{xcolor}
\usepackage{amsmath, amsthm, amsfonts, amssymb, epsfig}
\usepackage{tikz}
\usetikzlibrary{arrows,shapes}
\usepackage{caption}
\usepackage{subcaption}

\newcommand{\Z}{\ensuremath{\mathbb Z}}
\newcommand{\kP}{$k$-class}
\newcommand{\kPs}{$k$-classes}
\newcommand{\cM}{\mathcal{M}}

\newcommand{\prob}{P}

\newcommand{\m}{\cM}
\newcommand{\p}{\mathcal{P}}
\newcommand{\C}{C} 
\newcommand{\tn}{\tau_{nn}} 
\newcommand{\ttk}{\tau_{tk}}
\newcommand{\ttr}{\tau_{tree}}
\newcommand{\te}{\tau_{e}}
\newcommand{\ts}{\tau_{s}}
\newcommand{\tp}{\tau_{pp}} 
\newcommand{\mn}{\cM_{nn}}
\newcommand{\mpp}{\cM_{pp}}
\newcommand{\mk}{\cM_{k}} 
\newcommand{\mtk}{\cM_{tk}}
\newcommand{\mtr}{\cM_{tree}}
\newcommand{\me}{\cM_{e}}
\newcommand{\ms}{\cM_{s}}
\newcommand{\pk}{P_k} 
\newcommand{\pn}{P_n} 
\newcommand{\pmin}{p_{\text{min}}} 

\newcommand{\comment}[1]{}
\newcommand{\myInd}{\hspace*{1em}}
\usepackage{ifthen} 
\newboolean{includefigs} 
\setboolean{includefigs}{true} 
\newcommand{\condcomment}[2]{\ifthenelse{#1}{#2}{}}
\theoremstyle{plain}
\newtheorem{theorem}{Theorem}

\newtheorem{Definition}{Definition}
\newtheorem{Lemma}[theorem]{Lemma}
\newtheorem{Corollary}[theorem]{Corollary}

\title{Rapid Mixing of $k$-Class Biased Permutations}

\author{
  Sarah Miracle
    \thanks{Computer and Information Sciences, University of St.\ Thomas, St.\ Paul, MN 55105; {\tt sarah.miracle@stthomas.edu}.}
  \and Amanda Pascoe Streib
    \thanks{Center for Computing Sciences, Bowie, MD
      20715-4300; {\tt ampasco@super.org}.}}
\begin{document}
\date{}
\maketitle

\thispagestyle{empty}
\begin{abstract}
In this paper, we study a biased version of the nearest-neighbor
transposition Markov chain on the set of permutations where
neighboring elements $i$ and $j$ are placed in order $(i,j)$ with
probability $p_{i,j}$.   
Our goal is to identify the class of parameter sets ${\bf P} = \{p_{i,j}\}$ 
for which this Markov chain is rapidly
mixing. Specifically, we consider the open conjecture of Jim Fill that
all monotone, positively biased distributions are rapidly mixing.  

We resolve Fill's conjecture in the affirmative for distributions
arising from $k$-class particle processes, where the elements are divided into $k$ classes and the probability of exchanging neighboring elements depends on the particular classes the elements are in.  We further require that $k$ is a constant, and
all probabilities between elements in different classes are bounded
away from $1/2$.  
These particle processes arise in the
context of self-organizing lists and our result also applies beyond
permutations to the setting where all particles in a class are
indistinguishable.  Additionally we show that a broader class of
distributions based on trees is also rapidly mixing,  which
generalizes a class analyzed by Bhakta et. al. (SODA '13).  Our work generalizes recent work by Haddadan and Winkler (STACS '17) studying 3-class particle processes.
 
Our proof involves analyzing a generalized biased exclusion process, which is a nearest-neighbor transposition chain applied to a 2-particle system.  Biased exclusion processes are of independent interest, with applications in self-assembly.  
We generalize the results of Greenberg et al. (SODA '09) and Benjamini et. al (Trans.\ AMS '05) on biased exclusion processes 
to allow the probability of swapping neighboring elements to depend on the entire system, as long as the minimum bias is bounded away from $1$.
\end{abstract}
\newpage
\setcounter{page}{1}

\section{Introduction}
 The fundamental problem of generating a random permutation has a long history in
 computer science, beginning as early as 1969~\cite{Knuth2}.   
 One way to generate a random permutation 
 is to use the nearest-neighbor Markov chain $\mn$ which repeatedly swaps the elements in a random pair of adjacent positions.  The chain $\mn$ was among the first considered in the study of the
 computational  
 efficiency of Markov chains for sampling~\cite{ald,diasha,DSC93b} and
 has subsequently been studied extensively.  
After a series of papers, the mixing time  
of $\mn$ ($\Theta(n^3\log n)$~\cite{wilson}) is now well-understood when sampling from the uniform distribution on the permutation group~$S_n.$  

The nearest-neighbor chain $\mn$ can also be used to  
sample from more general
probability distributions by allowing non-uniform swap probabilities. 
Suppose we have a set of parameters ${\p}
=\{p_{i,j}\}$ and that $\mn$ puts neighboring elements $i$ and $j$ in
order $(i,j)$ with probability $p_{i,j}$, where $p_{j,i}=1-p_{i,j}$.  Despite the simplicity of this natural extension, much less is known about the mixing time of $\mn$ in the non-uniform case.
In this paper we look at the question for which parameter sets ${\p}$ is $\mn$ rapidly (polynomially)
mixing?  We say ${\p}$ is \emph{positively biased} if $p_{i,j}\geq
1/2$ for all $i<j$.  Without this condition, it is fairly straightforward to construct
parameter sets for which $\mn$ has mixing time that is exponential in
$n$ (see e.g.,~\cite{bmrs}).  Interestingly, Bhakta et al.~\cite{bmrs} showed that $\mn$ can require
exponential time to mix even for distributions with positive bias.   
In a widely circulated manuscript, Fill~\cite{F03a, F03b} introduced the following monotonicity conditions:  $p_{i,j} \leq p_{i,j+1}$ and $p_{i,j} \geq p_{i+1,j}$ for all $1 \leq i < j \leq n.$  Fill conjectured that $\mn$ is rapidly mixing for all monotone, positively biased distributions and further conjectured that the smallest spectral gap for $\mn$ among all of these distributions is given by the uniform $p_{i,j}=1/2$ distribution.  He confirmed these conjectures for $n\leq 3$ and gave experimental evidence for $n\leq 5$.

Except for a few special classes of monotone, positively biased distributions very little is known about the mixing time of $\mn$ in the non-uniform setting and the conjecture has remained unproven for over a decade.
Benjamini et al.~\cite{BBHM05} studied the case that $p_{i,j} = p>1/2$ for all $i<j$.  They showed that the mixing time of $\mn$ is $\Theta(n^2)$ for these distributions.  Bhakta et al.~\cite{bmrs} showed that if $p_{i,j}$ depends on only the smaller of $i$ and $j$, then $\mn$ mixes in polynomial time.  They extend this to distributions arising from binary trees with leaves labeled $\{1,2,\ldots, n\}$ and internal nodes labeled by probabilities, where the label of the lowest common ancestor of $i$ and $j$ in the tree determines~$p_{i,j}$.

Recently there has been interest in a special class of monotone,
positively biased distributions highlighted by Fill~\cite{F03b}.
These distributions are motivated by self-organizing lists and closely
related to exclusion processes arising in statistical physics.  In
this setting the set $[n]$ is partitioned into $k$ classes $\C_1, \C_2,
\ldots, \C_k$ and the probability of swapping two elements in $[n]$ is
determined by the classes containing those elements.
Define a \kP\ as a set of probabilities $\p$ where
elements from the same set are exchanged with probability 1/2 while
each element from $\C_i$ and each element from $\C_j$ (with $i < j$)
are put in increasing order with the same probability $p_{i,j} > 1/2.$
The setting of \kPs\ is motivated by self-organizing
lists, which have been in the literature for over 50 years  
 (see~\cite{HH85} for a survey). 
 Consider a set of records in a linear array, where element $i$ is
 requested with some unknown frequency $w_i$.  A self-organizing list
 is a way to reduce the linear look-up time by adjusting the
 permutation each time an element is requested.  The Move-Ahead-One
 (also called Transpose) algorithm updates the permutation by moving
 the element forward one position; i.e.\ it is precisely the nearest
 neighbor transposition chain.  The probability of swapping $i$ and
 $j$ is $p_{i,j} = w_i/(w_i+w_j)$.  We call this set of distributions
 $w$-distributions~\cite{F03b}, or more explicitly $k$-value
 $w$-distributions (where there are $k$ distinct frequencies $w_i$).
 Note these are an example of \kPs.

While $w$-distributions are quite natural and appear to be rapidly
mixing~\cite{F03b}, this particular simple instance of the biased permutation problem  
has so far eluded a thorough analysis.  Haddadan and
Winkler~\cite{HW16} recently studied 3-value $w$-distributions. 
They showed if $w_2/w_3, w_1/w_2 \geq 2$, then $\mn$ has mixing time
$O(n^{18}).$  They also analyzed a related nearest neighbor chain $\mpp$ over
$3$-particle systems, where the elements within a class are
indistinguishable, and so they are never swapped.   
They showed the mixing time of this chain is at most $O(n^{10}).$

\textbf{\underline{Our Main Result}.  }In this paper, we consider bounded \kPs, where $p_{i,j}/ p_{j,i}\geq
\gamma$ for all $i<j$ for some constant $\gamma>1$.  We show that if $\p$ is a weakly monotone
bounded \kP\ then the mixing time of $\mn$ is
$O(n^{2k+6}\log k)$.  This gives a polynomial bound for any constant
$k$, and applies directly to all bounded $k$-value
$w$-distributions (i.e. $w_i/w_{i+1}>\gamma$ for all $i$).  This improves the
mixing time bound given in~\cite{HW16} for $k=3$.  We also analyze
$\mpp$ over $k$-particle systems, and find the mixing time is $O(n^{2k+4})$, matching the
bounds from~\cite{HW16} for $k=3$.  In both cases, we extend their results to allow $\gamma< 2$.
In addition, we extend the work of Bhakta et
al.~\cite{bmrs} on distributions based on binary trees to
include trees with maximum degree at most~$k$.  

\textbf{\underline{Biased Exclusion Processes}.  }Simple 2-class particle systems, known as biased exclusion processes, have been a key tool in the study of biased permutations.  Suppose there are two types of particles (say 1 and 0) on a line, with $n_i$ (indistinguishable) particles of type $i$.  Define a (finite) biased exclusion process over the linear arrangements of these particles as follows: at each step, a pair of neighboring particles of different types may swap into increasing order with probability $p$ or out of order with probability $1-p$ .  Much of the previous work on the biased permutation problem 
has proceeded by mapping $\mn$ over permutations to several biased
exclusion processes or the related infinite asymmetric simple
exclusion processes (ASEPs).  
In~\cite{BBHM05}, $\mn$ is analyzed as a cross-product of several
ASEPs, and then rapid mixing for $\mn$ is inferred from the mixing
times of the ASEPs. 
Bhakta et al.~\cite{bmrs} discovered a different decomposition of
permutations into a cross-product of biased exclusion processes, which
allowed them to prove rapid mixing for more general $\p$
distributions.   

Exclusion processes are of independent interest, arising in a variety
of contexts.  The infinite version known as the asymmetric simple
exclusion process is a fundamental stochastic model in statistical
mechanics~\cite{Spitzer1970,BBHM05}.  
In combinatorics, the unbiased exclusion process is known as  
the mountain/valley Markov chain over monotonic lattice paths
(i.e. staircase walks)~(see Figure~\ref{staircase} and,
e.g.~\cite{LRS}).  Notice each linear arrangement of 1's and 0's can
be mapped bijectively to a lattice path in $\Z^2$ by sending 1's to
steps down and 0's to steps to the right.  A biased version of this
chain has applications in self-assembly, where it represents
reversible growth processes~\cite{GPR09, PR09}.

Benjamini et al.~\cite{BBHM05} bounded the mixing time of the
asymmetric exclusion process, where particles of type 0 and type 1 all
interact with the same (constant) probability $p$.  Subsequently,
Greenberg et al.~\cite{GPR09} discovered a simpler proof.  This
continues to be an active area of interest and in recent work  Labb\'e
et al.~\cite{LabbeLacoin} determined the exact mixing rate and Levin
et al.~\cite{LevinPeres} analyzed the case that $p$ tends to 0 as
$n\rightarrow\infty$.  Greenberg and others~\cite{PR09,GPRjournal} considered a heterogeneous biased exclusion process, 
 where the probability of swapping a 1 with a 0 at positions $i$ and
 $i+1$ depends on the number of 1's and the number of 0's to the left
 of position $i$.  
 
 In this paper, we introduce a new \emph{generalized
   (biased) exclusion process}, where the probability of swapping a 1
 with a 0 may depend on the entire sequence of 0's and 1's and prove it is rapidly mixing whenever the minimum bias is at least a constant.  Analyzing
 these processes is a key step towards proving our main result on
 permutations, and we believe it could be of interest beyond the
 application to biased permutations.  

\textbf{\underline{Techniques}.  }
In order to analyze $\mn$ we introduce a new Markov chain $\mtk$ which
includes a carefully selected set of more general transpositions
(swaps between non-nearest neighbor pairs).  The new chain allows us
to effectively break the single chain $\mn$ into a combination of
multiple unbiased permutation processes (one for each of the $k$
classes) and a single biased $k$-particle system where elements in the
same class are indistinguishable.  The bulk of our work is in proving
that the particle system is rapidly mixing.  Here our argument relies
on a novel decomposition argument where we  fix the location of all of
the particles in a single class and repeat this process inductively.
By doing this, we can again simplify permutations to several 2-class
particle systems as in previous work (\cite{BBHM05}, \cite{bmrs}),
with two key differences.  First, we reduce to our new generalized
biased exclusion processes mentioned above, where the probability of
swapping two particles depends on the entire state of the system.
Second, we need to use a decomposition theorem~\cite{mr} since in
general, monotone positively biased distributions do not appear to be
a simple cross-product of a set of 2-class particle systems. In fact,
we use decomposition inductively $O(k)$ times.

\section{The Markov Chains $\mn$ and $\mtk$.}\label{sec::prelim}
We begin by formalizing the Markov chain $\mn$.    
Then we will formally define a \kP\ and introduce an auxiliary chain $\mtk$ that  allows a larger set of transpositions.    
Let $\Omega = S_n$ be the set of all permutations $\sigma = (\sigma(1), \ldots, \sigma(n))$ of $n$ integers.  Suppose $\p$ is a set of probabilities, consisting of $p_{i,j} \in [0,1]$ for each $1 \leq i \neq j \leq n,$ where $p_{j,i} = 1- p_{i,j}.$  

\vspace{.1in}
\noindent  {\bf The Nearest Neighbor Markov chain $\mn$ } 

\vspace{.05in}
\noindent {\tt Starting at any permutation $\sigma_0$, iterate the following:} 
\begin{itemize}
\item At time $t,$ choose a position $1< i\leq n$ uniformly at random.  
\item With probability $p_{\sigma_t(i), \sigma_t(i-1)}$, exchange the elements $\sigma_t(i)$ and $\sigma_t(i-1)$ to obtain $\sigma_{t+1}$.
\item Otherwise, do nothing so that $\sigma_{t+1} = \sigma_{t}.$
\end{itemize}

\noindent The chain $\mn$ connects the state space $\Omega$ and has the stationary distribution (see e.g.,~\cite{bmrs})
$\pi(\sigma) = \left(\prod_{i<j}p_{\sigma(i), \sigma(j)}\right)Z^{-1},$ where $Z$ is the normalizing constant $\sum_{\sigma \in \Omega}\prod_{i<j}p_{\sigma(i), \sigma(j)}.$

For our main result we prove that if a set of probabilities $\p$ are weakly monotonic and form a bounded \kP\ then the Markov chain $\mn$ is rapidly mixing.  We will require the weakly monotonic condition defined in~\cite{bmrs} rather than the stronger monotonic condition defined in~\cite{F03b, F03a}. 

\begin{Definition}[\cite{bmrs}]\label{mono}The set $\p$ is weakly monotonic if properties 1 and either 2 or 3 are satisfied.
\begin{enumerate}
\item $p_{i,j} \geq 1/2$ for all $2 \leq i < j \leq n,$ and
\item $p_{i,j+1} \geq p_{i,j}$ for all $1 \leq i < j \leq n-1$ or
\item $p_{i-1,j} \geq p_{i,j}$ for all $2 \leq i < j \leq n.$
\end{enumerate}
\end{Definition}
\noindent For simplicity throughout the paper we will assume that property (2) holds.  If instead property (3) holds, the proofs are very similar and we point out distinctions throughout the paper.

Suppose $[n]$ is partitioned into $k$ particle classes  $\C_1, \C_2, \ldots, \C_k$.  Then a set of probabilities forms a \kP\ if particles in the same class interact with probability $1/2$ and the probability of swapping a particle in class $C_i$ with a (neighboring) particle in class $C_j$ is the same for all particles within those classes\footnote{ We assume, in order to ensure the distribution is positively biased, that $\C_1 = \{1,2,\ldots,c_1\}, \C_2 = \{c_1+1, \ldots, c_2\}, \ldots, \C_k = \{c_{k-1} + 1,\ldots,n\}$ for some $c_1 < c_2< \ldots < c_{k-1}$.}:
for all $ 1\leq i< j \leq k,$ if $x_1,x_2\in C_i$ and $y\in \C_j$ we have $p_{x_1,y} = p_{x_2,y}.$  We associate a permutation $\sigma$ with a \emph{$k$-particle system} where particles within the same class are indistinguishable.  For ease of notation, we record only the subscripts of the corresponding $C_i$ classes; that is, if $\sigma(1),\sigma(2)\in C_2, \sigma(3)\in C_1, $ and $\sigma(4),\sigma(5)\in C_3$, we will write 22133 for the associated $k$-particle system.
For any element $x$, let $\C(x)$ denote the particle class that
contains~$x$ (i.e. $\C(x) = i$ if and only if $x \in \C_i$).  Let
$\C_i > \C_j$ if $i>j$ and similarly for $\C_i = \C_j$.  We say a
\kP\ is \emph{bounded} if there exists a constant $\gamma>1$ such
that for all $1 \leq i < j \leq n,$ if $\C(i) \neq \C(j)$ then
$p_{i,j}/p_{j,i}\geq \gamma$.

Next we define a non-nearest neighbor Markov chain $\mtk$: $\mtk$ exchanges elements $\sigma(i)$ and $\sigma(j)$ at locations $i$ and $j$ with $i<j$ if for all $i<m<j,$  $\C(\sigma(m)) < \min(\C(\sigma(i)), \C(\sigma(j)))$.  That is, $\mtk$ swaps elements in different particle classes across elements in particle classes that are smaller than both.    
Particles in the same class can also be exchanged across any particles in other classes.  For example, suppose you start from a permutation with associated
$k$-particle system 3152673.  The chain $\mtk$ would allow an exchange
of the two 3 particles because they are in the same class and there
are no other particles in class 3 between them.  An exchange of the
first 3 and 5 (resulting in 5132673) would be allowed but an exchange
of 5 and 7 would not because there is an element in class 6 between
them.  Let $\lambda_{i,j} = p_{i,j}/p_{j,i}.$

\vspace{.1in}
\noindent  {\bf The Transposition Markov chain $\mtk$} 

\vspace{.05in}
\noindent {\tt Starting at any permutation $\sigma_0$, iterate the following:} 
\begin{itemize}
\item At time $t,$ choose a position $1\leq i \leq n$ and direction $d \in \{L, R,N\}$ uniformly at random.  
\item If $d = L$, find the largest $j$ with $1 \leq j<i$ and $\C(\sigma_t(j)) \geq \C(\sigma_t(i))$ (if one exists).  If $\C(\sigma_t(j)) > \C(\sigma_t(i)),$ then 
exchange the elements $\sigma_t(i)$ and $\sigma_t(j)$ to obtain $\sigma_{t+1}.$  \item If $d = R$, find the smallest $j$ with $n\geq j>i$ and $\C(\sigma_t(j)) \geq \C(\sigma_t(i))$ (if one exists).  If $\C(\sigma_t(j)) > \C(\sigma_t(i)),$ then with probability $\lambda_{\sigma_t(j),\sigma_t(i)}\prod_{i<k<j}\left(\lambda_{\sigma_t(j),\sigma_t(k)}\lambda_{\sigma_t(k),\sigma_t(i)}\right),$ exchange the elements $\sigma_t(i)$ and $\sigma_t(j)$ to obtain $\sigma_{t+1}.$  
\item If $d = N,$  find the largest $j$ with $1\leq j<i$ and $\C(\sigma_t(j)) = \C(\sigma_t(i)).$  If such an element exists, then exchange the elements $\sigma_t(i)$ and $\sigma_t(j)$ to obtain $\sigma_{t+1}.$ 

\item Otherwise, do nothing so that $\sigma_{t+1} = \sigma_t.$
\end{itemize}

\noindent We prove that $\mtk$ samples from the same distribution $\pi$ as $\mn$ (defined above) in Appendix~\ref{appendix::sameDist}.  

The time a Markov chain takes to converge to its stationary distribution, or \emph{mixing time}, is measured in terms of the distance between the distribution at time~$t$ and the stationary distribution. The \emph{total variation distance} at time~$t$ is
$\|\prob^t,\pi\| _{tv} = \max_{x\in\Omega}\frac{1}{2}\sum_{y\in\Omega} |\prob^t(x,y)-\pi(y)|,$
where $\prob^t(x,y)$ is the $t$-step transition probability.  For all $\epsilon>0$, the \emph{mixing time} $\tau(\epsilon)$ of $\m$ is defined as
$\tau(\epsilon)=\min \{t: \|\prob^{t'},\pi \|_{tv}\leq \epsilon, \forall t' \geq t\}.$
We say that a Markov chain is \emph{rapidly mixing} if the mixing time is bounded above by a polynomial in $n$ and $\log(\epsilon^{-1})$, where~$n$ is the size of each configuration in~$\Omega$.  In the remainder of the paper we will prove that $\mtk$ and then $\mn$ are rapidly mixing if the input probabilities $\p$ are weakly monotonic and form a bounded \kP.

\section{Bounded Generalized Exclusion Processes Mix Rapidly}\label{sec::exclusion}
We begin by analyzing bounded generalized biased exclusion processes.  Assume 
$n_1$ particles of type 1 and $n_0$ particles of type 0 occupy $n_0 + n_1$ linear positions: $1, \ldots, n_0+n_1$.  Let $\Omega_e$ be the set of all distinct orderings of $n_1$ 1's and $n_0$ 0's.    
In this setting, the probabilities $p_{\sigma_t,i}$ depend on both the current ordering $\sigma_t$ and the elements being exchanged.  Consider the following chain on $\Omega_e.$

\vspace{.1in}
\noindent  {\bf The Generalized Exclusion Markov chain $\me$ } 

\vspace{.05in}
\noindent {\tt Starting at any configuration $\sigma_0$, iterate the following:} 
\begin{itemize}
\item At time $t,$ choose a position $1\leq i< n_0+n_1$ uniformly at random.  
\item If $\sigma_t(i) \neq \sigma_t(i+1),$ with probability $p_{\sigma_t,i}$ exchange elements $\sigma_t(i)$ and $\sigma_t(i+1)$ to obtain $\sigma_{t+1}$.
\item Otherwise, do nothing so that $\sigma_{t+1} = \sigma_{t}.$
\end{itemize}

\noindent We say that $\me$ is \emph{bounded} if there exists a
constant $\gamma>1$ such that for all $\sigma \in \me$, if
$\sigma(i)=1$ and $\sigma(i+1)=0$ and $\tau$ is obtained
from $\sigma$ by swapping elements $\sigma(i)$ and $\sigma(i+1)$, then
$p_{\sigma,i}/p_{\tau,i}\geq \gamma$.

\begin{figure}[h]
  \centering
   \begin{tikzpicture}[scale=.6]
     \draw [fill=lightgray,thin,lightgray] (0,0) rectangle (1,3);
     \draw [fill=lightgray,thin,lightgray] (1,0) rectangle (2,2);
     \draw [fill=lightgray,thin,lightgray] (2,0) rectangle (4,1);
              \draw [fill=gray,thin,gray] (2,1) rectangle (3,2);
     \draw [help lines] (0,0) grid (4,3);
     \draw [ultra thick] (0,3) -- (1,3) -- (1,2) -- (3,2) -- (3,1) -- (4,1) -- (4,0);
   \end{tikzpicture}  
\hspace{2em}
    \begin{tikzpicture}[scale=.6]
     \draw [fill=lightgray,thin,lightgray] (0,0) rectangle (1,3);
     \draw [fill=lightgray,thin,lightgray] (1,0) rectangle (2,2);
     \draw [fill=lightgray,thin,lightgray] (2,0) rectangle (4,1);
         \draw [fill=gray,thin,gray] (2,1) rectangle (3,2);
     \draw [help lines] (0,0) grid (4,3);
     \draw [ultra thick] (0,3) -- (1,3) -- (1,2) -- (2,2) -- (2,1) -- (4,1) -- (4,0);
   \end{tikzpicture}
\caption{Two staircase walks that differ by a mountain/valley move (the dark grey square). } \label{staircase}
\end{figure}
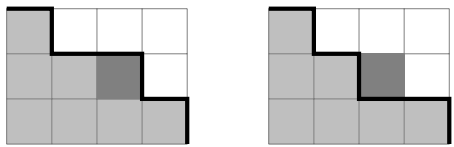

There is a straightforward bijection between $\Omega_e$ and staircase walks: map 1's to steps down and 0's to steps to the right.  For example, the two walks in Figure~\ref{staircase} map to 0100101 and 0101001 respectively. Exchanging a 1 and a 0 corresponds to adding or removing a particular square beneath the staircase walk.
 Greenberg and others~\cite{GPR09, PR09, GPRjournal} considered sampling monotonic surfaces in $\Z^2$ with bias.  
 They studied walks that start at $(0,h)$ and end at ($w,0$) and only
 move to the right or down and analyze a ``mountain / valley" chain
 that adds or removes a square along the boundary of the walk at each
 step (see Figure~\ref{staircase}).  In~\cite{GPRjournal}
 and~\cite{PR09}, they assumed the surface has ``fluctuating bias,"
 meaning that each square $s$ (on the $h\times w$ lattice) is assigned
 a bias $\lambda_s$ which is essentially the ratio of the
 probabilities of adding or removing that particular square.  They
 showed that as long as the minimum bias is a constant larger than 1
 then the chain is rapidly mixing.

  In our setting, the probability of adding or removing a particular square can vary depending on the rest of the configuration.  For example, the probability of moving from $1010$ to $1001$ is not the same as the probability of moving from 0110 to 0101. 
 We prove the following theorem.

\begin{theorem}\label{boundedBias} Let $\me$ be a bounded generalized exclusion process on $n_1$ 1's and $n_0$ 0's.  Suppose without loss of generality that $n_1 \leq n_0.$  Then the mixing time of $\me$ satisfies 
\abovedisplayskip=3pt
\belowdisplayskip=3pt
$$\tau(\epsilon) = O\left(\left(n_0+n_1\right)\left(n_1+\ln n_0+\ln \epsilon^{-1}\right)\right).$$
\end{theorem}

\noindent Our proof is similar to that of~\cite{GPRjournal} and we defer it to Appendix~\ref{sec::boundedBias}.  The idea is that the hitting time (time to reach  the most probable configuration) yields a bound on the mixing time, and if the minimum bias is a constant, then the hitting time is on the order of the area of the region. 

\section{$\mtk$ Mixes Rapidly for \kP\ Biased Permutations }\label{sec::kclass}
Next we prove that if the probabilities $\p$ form a \kP\ then the Markov chain $\mtk$ mixes rapidly. 
This will be useful when we analyze the nearest neighbor chain $\mn$ in Section~\ref{sec::kcomparison}.    
We first notice that the chain $\mtk$ is a product of $k+1$ independent Markov chains $\{\m_i\}$.  The first $k$ chains $\m_1, \m_2,\ldots, \m_k$ involve moves between particles in the same particle class and each such $\m_i$ is an unbiased nearest-neighbor Markov chain over permutations of $|C_i|$ particles.  The final chain $\m_{k+1}$ allows only moves between different particle classes.  The chains are defined formally in Appendix~\ref{appendix::subChains}.  
We prove the following theorem.
\begin{theorem}\label{mkMixing} If the probabilities $\p$ are weakly monotonic and form a bounded \kP\, then the mixing time $\ttk$ of the chain $\mtk$ satisfies $\ttk(\epsilon) = O\left(n^{2k}\ln(k\epsilon^{-1})\right)$, for $ k\geq 2$.
\end{theorem}
\noindent To prove Theorem~\ref{mkMixing}, we use a result of~\cite{bmrs} to relate the mixing times of the smaller chains $\{\m_i\}_{i=1}^{k+1}$ to $\mtk.$  Previous results~\cite{wilson} allow us to bound the mixing times of  $\m_1, \m_2,\ldots, \m_k$.  Thus, the bulk of our work is to bound the mixing time of~$\m_{k+1}$, which we do next.

\subsection{$k$-Particle Processes Mix Rapidly.}\label{sec::kparticle}  
Recall $\m_{k+1}$ allows only those moves of $\mtk$ that involve
elements in different particle classes (i.e. the moves with
direction $L$ and $R$).  
We call it a \emph{$k$-particle process} over its state space of $k$-particle systems,   since in this context 
 elements in the same class are indistinguishable.    
 If there are only two particle classes then this chain is a
bounded generalized exclusion process.    We
prove the following.  
   
\begin{Lemma}\label{pprocess}
Assume $|C_i|=c_i$ for all $i$.  The spectral gap $\text{Gap}(\prob_{k+1})$ of the chain $\m_{k+1}$ satisfies
\abovedisplayskip=3pt
\belowdisplayskip=3pt
\[\text{Gap}(\prob_{k+1}) =\Omega\Big(\Big(n^{k-1}\prod_{i=1}^{k-1}(c_i+\ln n)\Big)^{-1}\Big).\]
The mixing time of $\m_{k+1}$ satisfies $\tau_{k+1}(\epsilon)\leq O(n^{2k}\ln\epsilon^{-1}).$
\end{Lemma}

 Our proof will proceed inductively, and at each step of the induction
 we will apply the decomposition theorem~\cite{madr,mr00}.  We will
 use the following version of the decomposition theorem due to Martin
 and Randall~\cite{mr00}.  Let $\Omega = \cup_{i=1}^m \Omega_i$ be a
 partition of the state space into $m$ disjoint pieces.  For each
 $i=1,\ldots,m$, define $\prob_{i} = \prob(\Omega_i)$ as the
 restriction of $\prob$ to $\Omega_i$ which rejects moves that leave
 $\Omega_i$. In particular, the restriction to $\Omega_i$ is a Markov
 chain ${\cal M}_{i}$ with state space $\Omega_i$, where the
 transition matrix $\prob_{i}$ is defined as follows: If $x\not= y$
 and $x,y\in \Omega_i$ then $\prob_i(x,y)=\prob(x,y)$; if $x\in
 \Omega_i$ then $\prob_i(x,x)=1-\sum_{y\in \Omega_i, y\not= x}
 \prob_i(x,y)$.  Let $\pi_i$ be the normalized restriction of $\pi$ to
 $\Omega_i$, i.e., $\pi_i(A)=(\pi(A\cap\Omega_i))/(\pi(\Omega_i))$.
 Define $\widehat{\prob}$ to be the following aggregated transition
 matrix on the state space $\{1,\ldots,m\}$: 
$\widehat{\prob}(i,j)=\frac{1}{\pi(\Omega_i)}\sum_{\genfrac{}{}{0pt}{}{x\in\Omega_i,}{y\in
     \Omega_j}} \pi(x)\prob(x,y).$

\begin{theorem}[\cite{mr00}]\label{decomp}
Let $\prob_i=\prob(\Omega_i)$ and $\widehat{\prob}$ be as above, then
$\text{Gap}(\prob) \geq \frac{1}{2}\text{Gap}(\widehat{\prob})\min_{i=1,\ldots, m} \text{Gap}(\prob_{i}).$
\end{theorem}
\noindent We also use the following result to relate the mixing time to the spectral gap (see e.g., \cite{sinclair}, \cite{RM03}):
\begin{theorem}[\cite{RM03}]\label{gap}
Let $\pi_*=\min_{x\in\Omega} \pi (x)$.  For all $\epsilon>0$ we have\\
(a) $\qquad \tau(\epsilon)\geq \frac{|\lambda_1|}{2(1-|\lambda_1|)} \log\left(\frac{1}{2\epsilon}\right).$\\
(b) $\qquad \tau(\epsilon)\leq \frac{1}{1-|\lambda_1|} \log\left(\frac{1}{\pi_*\epsilon}\right).$
\end{theorem}

We may now prove Lemma~\ref{pprocess}.  As a running example, let
$\C_1 = \{1,2\}, \C_2=\{3\}, \C_3 = \{4,5\},$ and $\C_4 = \{6\}$.  Since elements within a class are indistinguishable to $\m_{k+1}$, we list each element using the subscript of its class; e.g., one 4-class particle system with these parameters is 412331. 

 \begin{proof}
For $i\geq 0$, let $\sigma_i$ represent an arbitrary fixed location of
the particles in classes $C_1,C_2,\ldots, C_i$ (when $i=0$, $\sigma_i$
represents no restriction).  For example,
$\sigma_{k-2} = \_12\_\_1$, where the $\_$ represents an empty
location which will be filled by an element of $\C_{k-1}$ or $\C_k$.
We will consider a smaller chain $\m_{\sigma_{i}}$ whose state space
is the set of all configurations where the elements in classes
$C_1,\ldots, C_i$ are in the locations given by $\sigma_i$: in our
example, the state space of $\m_{\sigma_{2}}$ is $\{{ \underline
  3}12{\underline{34}}1, {\underline 3}12{\underline{43}}1,
{\underline 4}12{\underline{33}}1\}$.  The moves of $\m_{\sigma_{i}}$
are a subset of the moves of $\mtk$.  It rejects all moves of $\mtk$
involving an element of $C_1,C_2,\ldots, C_i$.
We prove by induction that $\m_{\sigma_{i}}$ has spectral gap
satisfying
\abovedisplayskip=0pt
\belowdisplayskip=0pt
 \[\text{Gap}(\prob_i) = \Omega\Big(\Big(n^{k-1-i}\prod_{j=i+1}^{k-1}(c_j+
 \ln n)\Big)^{-1}\Big),\]
  for all choices of
 $\sigma_i$ (given a fixed $i$).  Since
 $\m_{\sigma_0}=\m_{k+1}$, this will prove the first part of Lemma~\ref{pprocess}.
  At each step of
 the induction, we apply decomposition (Theorem~\ref{decomp}).
 The restrictions of each decomposition will be rapidly mixing by
 induction and the projection chain will be a bounded generalized exclusion
 process.  The base case is $i=k-2$ and the final decomposition is $i=0$.

\textbf{Base case.} We begin with our base case, $i=k-2$.  Let
$\sigma_{k-2}$ be any fixed location of the particles in classes
$\C_1, \ldots, \C_{k-2}$.   
The Markov chain $\m_{\sigma_{k-2}}$ rejects all moves of $\mtk$ unless they
exchange a particle in class $\C_{k-1}$ with a particle in class
$\C_{k}$.  Thus, its moves only involve two types of particles, with
all other particles fixed, so we can view 
$\m_{\sigma_{k-2}}$ as a generalized exclusion process.  

Next we show that 
$\m_{\sigma_{k-2}}$ is bounded.  Consider any ``adjacent" particles $x\in \C_{k-1}$ and $y\in\C_{k}$ (they could be separated by any number of
particles in classes $\C_1, \ldots, \C_{k-2}$).  We select $x$ and
the appropriate direction (either $L$ or $R$) with probability
$1/(3n)$.  This succeeds with probability 1 if the direction is $L.$
If the direction is $R$, it succeeds with probability
$p_{k,k-1}/p_{k-1,k}$ if there are no additional 
particles between $x$ and $y$.  If there are
additional particles, then the probability is even smaller since we
are exchanging across elements in smaller classes and our
probabilities are weakly monotonic.  For example, moving from $312431$
to $412331$ happens with probability
$\frac{p_{4,3}p_{4,1}p_{4,2}p_{1,3}p_{2,3}}{p_{3,4}p_{1,4}p_{2,4}p_{3,1}
  p_{3,2}}\leq  \frac{p_{4,3}}{p_{3,4}}. $   Since for $i<j, \C(i)
\neq \C(j), p_{i,j} > 1/2,$ the minimum bias of our
generalized exclusion process $\m_{\sigma_{k-2}}$ satisfies  $\lambda_L = 1 / (p_{k,k-1}/p_{k-1,k}) =
p_{k-1,k}/(1-p_{k-1,k}) >1$.  Hence $\m_{\sigma_{k-2}}$ is bounded, so we can apply Theorem~\ref{boundedBias}.  
We have $c_{k-1}$
particles in class $\C_{k-1}$ and $c_{k}$ particles in class $\C_{k}$, and 
the moves of our exclusion process happen with probability $1/(3n)$
(instead of $1/(c_{k-1} + c_k)$). Thus, Theorem~\ref{boundedBias} (with $\epsilon=1/4$)
implies that for any such $\sigma_{k-2}$, $\m_{\sigma_{k-2}}$ has
mixing time $O(n\cdot\min(c_{k-1} + \ln c_k, c_k
+ \ln c_{k-1}) = O(n(c_{k-1}+\ln n)).$  Using Theorem~\ref{gap}(a)
we have that the spectral gap of $\m_{\sigma_{k-2}}$ satisfies
$\Omega(1/(n(c_{k-1} + \ln n)).$  

\textbf{Inductive Step.}   
We assume by induction the mixing time bound holds for all
$\m_{\sigma_i}$ for some $i\leq k-2$, and we
will use this result to prove that our mixing 
time bound holds for all $\m_{\sigma_{i-1}}$, which
fix the location of particles in one fewer particle class.  
Let $\sigma_{i-1}$ represent any fixed choice of locations for all
elements in classes $C_1,C_2,\ldots, C_{i-1}$.  In order to bound
$\text{Gap}(\prob_{i-1})$ we use the decomposition theorem.  Given any
$\sigma_i$ that is consistent with $\sigma_{i-1}$ (i.e. they agree on
the locations of all elements in classes $C_1,C_2,\ldots, C_{i-1}$),
the Markov chain $\m_{\sigma_i}$ will be a restriction Markov chain of
$\m_{\sigma_{i-1}}$, as defined in the decomposition theorem.  By
induction, we have $\text{Gap}(\prob_i)$  satisfies $\text{Gap}(\prob_i) =
\Omega\Big(\Big(n^{k-1-i}\prod_{j=i+1}^{k-1}(c_j+ \ln n)\Big)^{-1}\Big).$

The projection chain, however, is more complicated. Recall our running example ($k=4$) and consider the second decomposition.  Here each of the restrictions is the set of configurations consistent with a particular fixed location of the particles in classes $C_1$ and $C_2$ and all restrictions agree on the location of particles in $C_1.$  
Let $\sigma_2 =  \_112\_\_ $ and $\beta_2 =  211\_\_\_ $ represent two such restrictions.  A move of the projection chain between $\sigma_2$ and $\beta_2$ is an aggregate of all moves of $\mtk$ between configurations consistent with $\sigma_2$ and configurations consistent with $\beta_2.$  For example, $411233 \rightarrow 211433.$  Each of these moves involve exchanging a particle in $C_3$ or $C_4$ with a particle in $C_2$.  However, since these exchanges may happen across any number of particles in $C_1$ and involve particles in $C_3$ or $C_4$ they will have different probabilities, making the analysis more challenging.    

More generally, moves of the projection chain involve exchanging an element
from $\C_i$ with an element from $\C_j$ where $j >i$.  There may be
additional elements between the elements being exchanged but if there
are, they are in a smaller particle class $\C_s$ with $s < i.$  If we
view all elements in $\C_i$ as one type and all elements in $\C_{i+1}, \C_{i+2},
\ldots, \C_{k} $ as another, then the projection chain can be viewed as a
bounded generalized exclusion process.  Specifically, we will show that all moves
that move a particle in $\C_i$ ahead happen with probability $1/(3n)$
and all moves that move it back happen with probability at most
$(1/(3n))(p_{i+1,i}/p_{i,i+1}).$  Since $p_{i,i+1} = 1 - p_{i+1,i} >
1/2,$ this implies that the minimum bias is greater than 1 and we can
apply Theorem~\ref{boundedBias}.  There are $c_i$ particles of type
$i$, $\sum_{j=i+1}^k c_j < n$ particles of the other type, and the
moves are selected with probability $1/(3n).$   Applying
Theorem~\ref{boundedBias} and Theorem~\ref{gap}(a) shows that the
spectral gap of the projection chain satisfies $\Omega((n(c_i + \ln n) )^{-1})$.    Combining this with the bound on the
restriction chain, Theorem~\ref{decomp} implies 
\begin{equation}\label{gapeq}
\text{Gap}(\prob_{i-1})= \Omega\Big(\Big(n^{k-i}\prod_{j=i}^{k-1}(c_j+\ln n)\Big)^{-1}\Big).  
\end{equation}

It remains to show that the projection chain
$\widehat{\m}_{\sigma_{i-1}}$ moves an element in $\C_{i}$ backward
with probability at most $(1/(3n))(p_{i+1,i}/p_{i,i+1}).$  Without
loss of generality, consider a move $(\gamma, \beta)$ of the
projection chain $\widehat{\m}_{\sigma_i}$, which exchanges an element
in $\C_i$ at location $a$ with an element at location $b$ where $b >
a.$  Note that it is possible that $b \neq a+1$ but if that is the
case then for all $b > c > a, \p(c) < i.$  The definition of the
projection chain from Theorem~\ref{decomp}  gives us that \[
\widehat{\prob} (\gamma, \beta) =
\frac{1}{\pi(\Omega_{\gamma})}\sum_{x \in \Omega_\gamma, y \in
  \Omega_\beta}\pi(x)\prob(x,y).\]
Recall that $\Omega_{\gamma}$ consists of all configurations that have
the elements in particles classes $C_1, C_2,\ldots, C_i$ fixed in the
identical locations as those in $\gamma$ and configurations in
$\Omega_{\beta}$ have the location of those elements in the same
position except that there is an element of type $\C_i$ in position
$b$ instead of position $a$.  Now consider any configuration $x \in
\Omega_{\gamma},$  if we select position $a$ and direction $R$ then we
will exchange the elements at position $a$ and $b$ with probability
$\frac{p_{j,i}}{p_{i,j}}\prod_{a<c<b}\left[\left(\frac{p_{j,\gamma(c)}}{p_{i,\gamma(c)}}\right)\left(\frac{p_{\gamma(c),i}}{p_{\gamma(c),j}}\right)\right]).$
Since $\gamma(c) < i < j,$ the weak monotonicity condition property 2
(Definition~\ref{mono})\footnote{If the input probabilities $\p$
  satisfy the weak monotonicity property 3 instead then we would need
  to modify $\mtk$ to instead allow swaps between elements in
  different particle classes across elements whose particle class is
  larger (instead of smaller).  This proof could then be easily
  modified so that the base case restricts the location of particles
  in $\p_3, \p_4,\ldots, \p_k,$ $\sigma_{i}$ represents a particular fixed
  location of all the particles in $\C_i, \C_{i+1}, \ldots, \C_k$ and so
  forth.}  implies that $p_{\gamma(c), i} \leq p_{\gamma(c), j}$ and
$p_{j,\gamma(c)} \leq p_{i,\gamma(c)}$ and thus
$\left[\left(\frac{p_{j,\gamma(c)}}{p_{i,\gamma(c)}}\right)\left(\frac{p_{\gamma(c),i}}{p_{\gamma(c),j}}\right)\right]
< 1$ for all such $c.$  Similarly, since $j \geq i+1$ we have that
$p_{i,j} > p_{i,i+1}$ and $p_{j,i} < p_{i+1,i}.$  Combining these
gives the following
$$\widehat{\prob} (\gamma, \beta) 
\leq \frac{1}{\pi(\Omega_{\gamma})}\sum_{x \in \Omega_\gamma}\pi(x)
\left(\frac{1}{3n}\right)\left( \frac{p_{i+1,i}}{p_{i,i+1}}\right)
=  \frac{1}{3n}\left(\frac{p_{i+1,i}}{p_{i,i+1}}\right).$$

Finally, we will bound $\tau_{k+1}(\epsilon)$ for $\epsilon>0$.  Let $\lambda_* = \max_{i<j} p_{i,j}/p_{j,i}$ then $\pi_* =  \min_{x\in \Omega}\pi(x) \geq (\lambda_*^{\binom{n}{2}}n!)^{-1}$ (see \cite{bmrs} for more details), so $\log(1/\epsilon\pi_*) = O(n^2\ln \epsilon^{-1})$ since $\lambda$ is bounded from above by a positive constant.  Applying Theorem~\ref{gap}(b) and (\ref{gapeq}), we have $\tau_{k+1}(\epsilon)\leq O(n^{2(k-1)}n^2\ln\epsilon^{-1})$.
\end{proof}

\subsection{From $k$ Particle Process to \kP.}\label{particleToClass}
Again, we can view $\mtk$ as a product of $k+1$ smaller Markov chains
where $\m_1, \m_2,\ldots, \m_k$ are unbiased nearest-neighbor chains
over permutations of a single particle class (moves between elements
in the same particle class) and $\m_{k+1}$ is a $k$-particle process
(moves between elements in different particle classes).  We will use
the following result of Wilson to bound the mixing times of  the $k$
permutation processes and Lemma~\ref{pprocess} to bound the
mixing time of $\m_{k+1}.$  

\begin{theorem}[\cite{wilson}]\label{unbiased}
The chain $\mn$ mixes in time $O(n^3\log n\log\epsilon^{-1})$ when $p_{ij} = 1/2$ for all $i<j.$
\end{theorem}
\noindent Let $\tau_i$ be the mixing time of $\m_i$ for $1\leq i\leq
k+1$.  Each chain $\m_{i}$ for $1 \leq i \leq k$ has $c_i$ particles, so Theorem~\ref{unbiased} implies $\tau_i(\epsilon) =O(c_i^3\log c_i\log\epsilon^{-1}).$ 
By Lemma~\ref{pprocess}, $\m_{k+1}$ has mixing time $O(n^{2k}\ln\epsilon^{-1})$. 
To bound the mixing time of $\mk,$ we will use the following
theorem due to Bhakta et al.~\cite{bmrs}, which bounds the mixing time
of a product of independent Markov chains.

\begin{theorem}[\cite{bmrs}]\label{prodID}
Suppose the Markov chain $\m$ is a product of $N$ independent Markov chains $\{\m_i\},$ where $\m$ updates $\m_i$ with probability $p_i.$  If $\tau_i(\epsilon)$ is the mixing time for $\m_i$ and $\tau_i(\epsilon)\geq 4\ln \epsilon$ for each $i,$ then $\tau(\epsilon)\leq \max_{i=1,2,\ldots,N}\frac{2}{p_i}\tau_i\left(\frac{\epsilon}{2N}\right).$
\end{theorem}

\noindent The Markov chain $\mtk$ will update $\m_i$ for $1\leq i \leq k$ if direction $N$ is selected and a particle in class $\p_i$ which happens with probability $c_i/(3n).$  The Markov chain $\m_{k+1}$ is updated when direction $L$ or $R$ is selected; i.e. with probability $2/3$.  Therefore, for $i\leq k$, 
\[\frac{2}{p_i}\tau_i\left(\frac{\epsilon}{2(k+1)}\right) = O(nc_i^2\ln c_i\ln(k\epsilon^{-1})) = O(n^3\ln n \ln(k\epsilon^{-1})).\]
On the other hand, 
$\frac{2}{p_{k+1}}\tau_{k+1}\left(\epsilon/(2(k+1))\right) = O(n^{2k}\ln(k\epsilon^{-1})).$  Therefore, Theorem~\ref{prodID} gives that $\ttk(\epsilon) = O(n^{2k}\ln(k/\epsilon))$ for $k\geq 2$. This proves Theorem~\ref{mkMixing}.

\section{$\mn$ Mixes Rapidly for \kP\ Permutations.}\label{sec::kcomparison}

In the final part of our argument we will use the comparison method~\cite{dsc,RT98} to bound the mixing time of $\mn$ using the bound on the mixing time of $\mtk$ (Theorem~\ref{mkMixing}).  We will prove the following.
\begin{theorem}\label{mnMixing} If the probabilities $\p$ are weakly monotonic and form a bounded \kP\ with $k \geq 2$, then the mixing time $\tn$ of the chain $\mn$ satisfies 
$\tn(\epsilon) = O\left(n^{2k+6}\ln(k\epsilon^{-1}) \right).$
\end{theorem}
\noindent As a corollary we prove the following result on $k$-particle systems (see Section~\ref{appendix::particleprocess}).  In this setting, particles in the same class are indistinguishable.  The particle process chain $\mpp$ is identical to $\mn$ except exchanges are only allowed between elements in different classes.
\begin{Corollary}\label{mpMixing} If the probabilities $\p$ are weakly monotonic and form a bounded \kP\ with $k \geq 2$, then the mixing time $\tp$ of the chain $\mpp$ satisfies 
$\tp(\epsilon) = O\left(n^{2k+4}\ln\epsilon^{-1} \right).$
\end{Corollary}  
We will use the following form of the comparison method due to Randall and Tetali~\cite{RT98}.
Let $P'$ and $P$ be two reversible Markov chains on the same state space $\Omega$ with the same stationary distribution $\pi$ and let $E(P) = \{(x,y): P(x,y) > 0\}$ and $E(P') = \{(x,y): P'(x,y) > 0\}$ denote the sets of edges of the two graphs, viewed as directed graphs.  For each $x,y$ with $P'(x,y)>0$, define a path $\gamma_{xy}$ using a sequence of states $x=x_0,x_1,\cdots,x_k = y$ with $P(x_i,x_{i+1})>0$, and let $|\gamma_{xy}|$ denote the length of the path.  Let $\Gamma(z,w) = \{(x,y) \in E(P'): (z,w) \in \gamma_{xy}\}$ be the set of paths that use the transition $(z,w)$ of $P$.  Finally, define 
\abovedisplayskip=3pt
\belowdisplayskip=3pt
$$A = \max_{(z,w) \in E(P)} \left \{\frac{1}{\pi(z)P(z,w)}\sum_{\Gamma(z,w)}|\gamma_{xy}|\pi(x)P'(x,y) \right \}.$$

\begin{theorem}[\cite{RT98}]\label{comparison}Given two Markov chains each with stationary distribution $\pi,$ transition matrices $P$ and $P'$ and mixing times $\tau(\epsilon)$ and $\tau'(\epsilon),$ respectively.  Define $A$ and $\pi_*$ as above, then for $0<\epsilon <1,$ we have
$\tau(\epsilon) \leq 4\log(1/(\epsilon\pi_*))A\tau'(\epsilon)/\log(1/2\epsilon).$
\end{theorem}

We can use the same bound on the minimum weight configuration $\pi_*$, $\log(1/\pi_*) = O(n^2)$ given in Section~\ref{sec::kparticle}, so it remains to bound the quantity $A.$  For each edge $(x,y)$ in $\mtk$ we will define a path $\gamma_{xy}$ using edges in $\mn$.  In each step of $\mtk$ we choose a position and then a direction.  We will define different paths based on the direction.  The paths for direction $N$ are the most complex and we provide a high-level description here.  The full description of the three types of paths (corresponding to the three directions) and the complete details of the proofs can be found in Section~\ref{appendix::comparison}.  
Suppose $x$ and $y$ differ by an $(a,b)$ transposition, where $\C(a) = \C(b)$ and $a$ and $b$ are not adjacent. 
Roughly, our path will move element $b$ to the left until it reaches the correct location and then move element $a$ to the right.  However, we must design our paths very carefully to ensure that the weight of the intermediate configurations is never less than $\min (\pi(x), \pi(y)).$  For example, if the configuration is 38881113 (the numbers refer to the particle classes the elements reside in) and our goal is to exchange the two 3's, we cannot simply move the leftmost 3 to the right (or rightmost 3 to left) because this will decrease the weight.  

In the first phase, our path will move $b$ to the left until we reach an element in a smaller particle class than $b.$  Let $c$ be the first element to the left of $b$ such that $\C(c)>\C(b)$, and suppose $c$ is in position $i$.  Notice swapping $c$ and $b$ would not decrease the weight of the configuration.  However, this is not a nearest-neighbor swap.  To execute this swap, move $c$ to the right until it passes $b$ (by construction, none of these moves decreases the weight).  Then move $b$ to the left to position $i$.  Repeat this process (with new choices of the intermediate particle $c$) until,
eventually, $a$ and $b$ are adjacent.  At this point, swap $a$ and $b$.  In the second phase of the path, element $a$ will retrace the same exact steps that $b$ took originally (including replacing any elements in larger particle classes that were moved).  For an example path, see Figure~\ref{kpath}.  The paths are formally described in Figure~\ref{canonicalPaths} in the appendix.  Note that for direction $N$ our paths use similar ideas to those in~\cite{bmrs} (Section 5.2), but this version gives a better bound since we are swapping elements in the same particle class. 

\begin{figure}
\begin{tabular}{ l l }
Step \# & Configuration\\
\hline
0 & {\bf $3_a$}\ 5\ 12\ 86\textcolor{red}{7}\ 121\ \bf $3_b$\\
1 & {\bf $3_a$}\ 5\ 12\ 86\ 121\ {\bf $3_b$}\textcolor{red}{\ 7}\\
2 &  {\bf $3_a$}\ \textcolor{red}{5}\ 12\ {\bf $3_b$}86\ 121\ 7\\
3 &   {\bf $3_a$}\ 12\ {\bf $3_b$}\textcolor{red}{5}86\ 121\ 7\\
4 & {\bf $3_a$} {\bf $3_b$}\ 12\ 586\ 121\ 7\\
\end{tabular}
\hspace{2em}
\begin{tabular}{ l l }
Step \# & Configuration\\
\hline
5 &  {\bf $3_b$} {\bf $3_a$}\ 12\ 586\ 121\ 7\\
6 &   {\bf $3_b$}\ 12\  {\bf $3_a$}\textcolor{red}{5}86\ 121\ 7\\
7 &  {\bf $3_b$}\ \textcolor{red}{5}\ 12\  {\bf $3_a$}86\ 121\ 7\\
8 &  {\bf $3_b$}\ 5\ 12\  86\ 121\ {\bf $3_a$}\textcolor{red}{7}\\
9 & {\bf $3_b$}\ 5\ 12\  86\textcolor{red}{7}\ 121\ {\bf $3_a$}\\
\end{tabular}
\caption{The critical steps in $\gamma_{x,y}$ for $x = 3_a565212713_b$ and $y= 3_b565212713_a$ (the numbers refer to the particle classes of the elements between $3_a$ and $3_b$ which are both in particle class 3). }\label{kpath}
\vspace{-.5em}
\end{figure}


\section{Trees of $k$-Value Permutations Mix Rapidly}\label{sec::tree}

Bhakta et al~\cite{bmrs} define a class of probabilities they call ``League Hierarchies" and show that this class mixes rapidly.  A set of probabilities $\p$ is in this class if there exists a binary tree $T$ with $n$ leaves labeled $1,\ldots, n$ in sorted order where each non-leaf node $v$ has a value $\frac{1}{2} \leq q_v < 1$ associated with it and $p_{i,j} = q_{i\wedge j}$ where $i \wedge j$ is the lowest common ancestor of the leaves labeled $i$ and $j$ in $T$.  At a high-level, they show that if the probabilities $\p$ have this type of structure, then we can view the chain as a collection of independent biased 2-particle exclusion processes.  In this section, we extend their league hierarchies beyond binary trees using our result that for constant $k$, $k$-particle processes are rapidly mixing.  
  We can now allow tree nodes with up to $k$ children.  

\begin{figure}[tp]
\centering
\subcaptionbox{\label{treePic}}{
\begin{tikzpicture}[>=,level/.style={sibling distance = 2.5cm,
  level distance = 2cm, inner sep = 0, minimum size = .5cm},scale=.5]
\node [circle,draw,inner sep = 0, minimum size = .5cm]  {A}
    child{ node [circle,draw]  {B} 
            child{ node [circle,draw,fill=lightgray!50]  {1} edge from parent node[left]{\tiny 1}
            }
            child{ node[circle,draw,fill=lightgray!50]  {2} edge from parent node[left]{\tiny 2}
            }
						child{ node[circle,draw,fill=lightgray!50]  {3} edge from parent node[left]{\tiny 3}
            }edge from parent node[left]{\tiny 1}
    }
    child{ node [circle,draw,fill=lightgray!50]  {4} edge from parent node[left]{\tiny 2}
		}
		child{ node [circle,draw] {C} 
            child{ node [circle,draw,fill=lightgray!50]  {5} edge from parent node[left]{\tiny 1}
            }
            child{ node [circle,draw,fill=lightgray!50]  {6} edge from parent node[left]{\tiny 2}                                          
            } edge from parent node[right]{\tiny 3}
                }
		child{ node [circle,draw,fill=lightgray!50]  {7} edge from parent node[right]{\tiny 4}
		}
	
;
\end{tikzpicture}
}
\subcaptionbox{\label{qValues}}{
\begingroup
\tiny
\begin{tabular}{ l  }
  $q_{(A,1,2)} = .6$ \\
  $q_{(A,1,3)} = .7$ \\
	$q_{(A,1,4)} = .8$ \\
	$q_{(A,2,3)} = .8$ \\
	$q_{(A,2,4)} = .8$ \\
	$q_{(A,3,4)} = .7$ \\
	$q_{(B,1,2)} = .6$ \\
	$q_{(B,1,3)} = .8$ \\
	$q_{(B,2,3)} = .8$ \\
	$q_{(C,1,2)} = .9$ \\
\end{tabular}
\endgroup
}
\subcaptionbox{\label{treepic2}}{
\begin{tikzpicture}[>=,level/.style={sibling distance = 2.5cm,
  level distance = 2cm, inner sep = 0, minimum size = .5cm},scale=.5]
\node [ellipse,draw, inner sep = 0, minimum size = .5cm]  {\tiny $\genfrac{}{}{0pt}{}{(6143275)}{\bf 3121143}$\normalsize}
    child{ node [ellipse,draw]  {\tiny $\genfrac{}{}{0pt}{}{(132)}{\bf 132}$\normalsize} 
            child{ node [circle,draw,fill=lightgray!50]  {1} edge from parent node[left]{\tiny 1}
            }
            child{ node[circle,draw,fill=lightgray!50]  {2} edge from parent node[left]{\tiny 2}
            }
						child{ node[circle,draw,fill=lightgray!50]  {3} edge from parent node[left]{\tiny 3}
            }edge from parent node[left]{\tiny 1}
    }
    child{ node [circle,draw,fill=lightgray!50]  {4} edge from parent node[left]{\tiny 2}
		}
		child{ node [ellipse,draw] {\tiny$\genfrac{}{}{0pt}{}{(65)}{\bf 21}$\normalsize} 
            child{ node [circle,draw,fill=lightgray!50]  {5} edge from parent node[left]{\tiny 1}
            }
            child{ node [circle,draw,fill=lightgray!50]  {6} edge from parent node[left]{\tiny 2}                                          
            } edge from parent node[right]{\tiny 3}
                }
		child{ node [circle,draw,fill=lightgray!50]  {7} edge from parent node[right]{\tiny 4}
		}
;
\end{tikzpicture}
}
\caption{
\small
A set of probabilities $\p$ with $k$-league structure, the corresponding $q$ values, and the tree representation of the permutation $6143275.$
}\label{bigTreePic}
\end{figure}
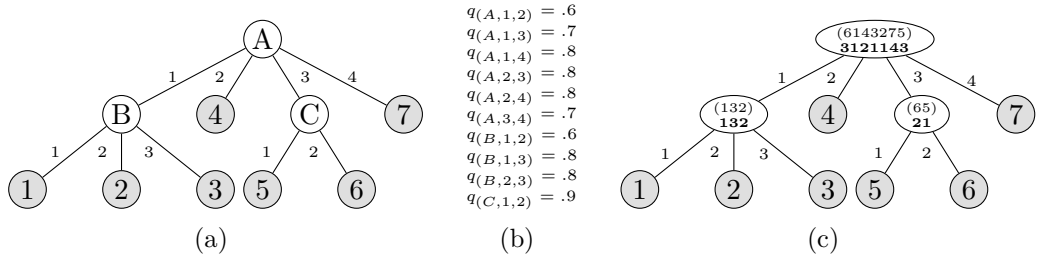

Let $T$ be a labeled ordered tree (or plane tree) with $n$ leaves labeled $1,\ldots, n$ in sorted order.  For each internal node $v$, the children of $v$ are labeled $1,2,\ldots, deg(v)$, and if  $deg(v) \geq 2$, $v$ is assigned $\binom{deg(v)}{2}$ values $1/2 < q_{(v,a,b)}<1,$ for $1\leq a < b \leq deg(v)$ (here $a$ and $b$ correspond to the labels of the children of $v$).   Again let $i \wedge j$ be the lowest common ancestor of the leaves labeled $i$ and $j.$  We say that a set of probabilities $\p$ has \emph{$k$-league structure} if there exists such a tree $T$ with $p_{i,j} = q_{(i\wedge j, a,b)}$ where $a$ and $b$ are children of $i \wedge j$, $i$ is contained in the subtree rooted at $a$ and $j$ is contained in the subtree rooted at~$b.$  Figure~\ref{treePic} gives an example $\p$ with $k$-league structure.  In this example, $p_{2,6} = q_{(A,1,3)} = .7$ since $2\wedge 6 = A$ and leaves $2$ and $6$ come from $A$'s subtrees labeled $1$ and $3$ respectively.  We prove the chain $\mn$ is rapidly mixing for any $\p$ with $k$-league structure.  

\begin{theorem} Given a set of probabilities $\p$ that is weakly monotonic with bounded $k$-league structure and corresponding tree $T$ with maximum degree $k,$  the mixing time $\tn$ of $\mn(T)$ satisfies $$\tn(\epsilon) =  O\left(n^{2k+11}\ln(n\epsilon^{-1})\right).$$
\end{theorem}

\noindent We will first prove that the tree chain $\mtr$, introduced by~\cite{bmrs} is rapidly mixing and then use the comparison theorem (Theorem~\ref{comparison}) to relate the mixing time of $\mtr$ to the mixing time of $\mn.$  Bhakta et al~\cite{bmrs} prove that $\mtr$ has the same stationary distribution as $\mn.$

\vspace{.1in}
\noindent {\bf  The Markov chain $\mtr$}

\vspace{.05in}
\noindent{\tt  Starting at any permutation $\sigma_0$, iterate the following:}

  \begin{itemize}
    \item Select distinct $a, b \in [n]$ with
       $a < b$ u.a.r.
    \item If every number between $a$ and $b$ in the permutation $\sigma_t$
      is not a descendant in $T$\\
      of $a\wedge b$, obtain $\sigma_{t+1}$ from
      $\sigma_t$ by placing $a,b$ in order with probability
      $p_{a,b}$, and out of order with probability $1 - p_{a,b}$, leaving all
      elements between them fixed.
    \item Otherwise, $\sigma_{t+1}=\sigma_t$.
  \end{itemize}

 As in~\cite{bmrs} we will decompose the chain $\mtr$ into at most
 $n-1$ independent Markov chains, one for each non-leaf node of the
 tree $T.$  In their case they introduce an alternative representation
 of a permutation with league structure as a tree of binary strings,
 one for every non-leaf node.  We will generalize this representation
 so that an internal node $v$ with $deg(v) = k$ will now hold a string
 of numbers from $\{1, \ldots, k\}.$  We will refer to this as the
 node's \emph{tree string}.  The number of $i$'s in the string will be
 the number of leaves in the subtree rooted at child $i$ of node $v.$
 Figure~\ref{qValues} gives an example of the tree representation of a
 particular permutation. We show that this representation (which we
 will call the \emph{tree representation}) is a bijection in
 Appendix~\ref{appendix::tree}.

 The permutation $\sigma$ is the permutation string of the root node.
 We will analyze $\mtr$ by considering its effect on the tree
 representation of a permutation.  Each move of $\mtr$ involves
 swapping two elements $a$ and $b$ which correspond to adjacent (and
 different) numbers in $a \wedge b$'s tree string.  Note that for any
 other ancestor of $a$ and $b$, both these nodes correspond to the
 same number in that node's tree string and thus the move does not
 change the string at that node.  For any descendent of $a\wedge b$
 the move also does not change the tree string because at most one of
 $a$ and $b$ is represented in that node's tree string and since $a$
 and $b$ are only exchanged across nodes that are not descendants of
 $a \wedge b$, any descendant's string remains the same.  Thus the $a$
 and $b$ exchange only modifies the tree string at node $a \wedge b$
 and thus the particle processes at each node are independent.  Given
 this, we can use our mixing time result for $k$-particle processes
 (Corollary~\ref{mpMixing}) combined with Theorem~\ref{prodID} which
 bounds the mixing time of a chain that is a product of independent
 Markov chains to bound the mixing time of $\mtr.$   Given a node
 whose tree string has length $b$ (i.e. the node has $b$ descendants
 that are leaves), the probability of selecting a move that
 corresponds to two neighboring characters in the string is
 $(b-1)/\binom{n}{2} = \frac{b-1}{2n(n-1)}.$  Let~$k$ be the maximum degree of~$T.$  Then 
 Corollary~\ref{mpMixing} and Theorem~\ref{prodID} imply
$$\ttr(\epsilon) = O\left(\frac{n(n-1)}{b-1}(b^{2k+4}\ln (2n/\epsilon)\right) = O\left(n^{2k+5}\ln(n\epsilon^{-1})\right).$$

Finally, to relate the mixing time of $\mtr$ to the mixing time of $\mn$ we can use the comparison theorem (Theorem~\ref{comparison}).  Since $\mtr$ will not exchange elements $a$ and $b$ across elements $c$ with $\C(a) < \C(c) < \C(b)$ (elements in particle classes between $a$ and $b$) we can use paths that are almost identical to those described in Section~\ref{sec::kcomparison} (the details can be found in Section~\ref{appendix::comparison})  for moves with direction $N.$  However since $\C(a)$ and $\C(b)$ may not be equal, in order to recover an edge of $\mtr$ from an edge of $\mn$ we will need to know the original location of both $\C(a)$ and $\C(b).$  This extra factor of $n$ is balanced out by the fact that moves of $\mtr$ are selected with probability $1/\binom{n}{2}$ while in $\mn$ they are selected with probability $1/n.$  Thus the bound on $A$ remains $O(n^4)$.  Using this bound on $A$, our bound on the minimum weight configuration $\pi_*$ from Section~\ref{particleToClass} and appealing to the comparison theorem (Theorem~\ref{comparison}) and our above bound on the mixing time of $\mtr$ gives 
$\tn(\epsilon)=  O\left(n^{2k+11}\ln(n\epsilon^{-1})\right).$

\bibliographystyle{plain}
\bibliography{permutationsbib}

\begin{thebibliography}{10}

\bibitem{ald}
D.~Aldous.
\newblock Random walk on finite groups and rapidly mixing {M}arkov chains.
\newblock {\em In Seminaire de Probabilites XVII}, pages 243--297, 1983.

\bibitem{BBHM05}
I.~Benjamini, N.~Berger, C.~Hoffman, and E.~Mossel.
\newblock Mixing times of the biased card shuffling and the asymmetric
  exclusion process.
\newblock {\em Trans. Amer. Math. Soc}, 357:3013--3029, 2005.

\bibitem{bmrs}
P.~Bhakta, S.~Miracle, D.~Randall, and A.P. Streib.
\newblock Mixing times of {M}arkov chains for self-organizing lists and biased
  permutations.
\newblock In {\em Proceedings of the 24th ACM/SIAM Symposium on Discrete
  Algorithms}, SODA '13, pages 1--15, 2013.

\bibitem{DSC93b}
P.~Diaconis and L.~Saloff-Coste.
\newblock Comparison techniques for random walks on finite groups.
\newblock {\em The Annals of Applied Probability}, 21:2131--2156, 1993.

\bibitem{dsc}
P.~Diaconis and L.~Saloff-Coste.
\newblock Comparison theorems for reversible {M}arkov chains.
\newblock {\em Annals of Applied Probability}, 3:696--730, 1993.

\bibitem{diasha}
P.~Diaconis and M.~Shahshahani.
\newblock Generating a random permutation with random transpositions.
\newblock {\em Z. Wahrscheinlichkeitstheorie Verw. Gebiete}, 57:159--179, 1981.

\bibitem{F03b}
J.~Fill.
\newblock Background on the gap problem.
\newblock {\em Unpublished manuscript}, 2003.

\bibitem{F03a}
J.~Fill.
\newblock An interesting spectral gap problem.
\newblock {\em Unpublished manuscript}, 2003.

\bibitem{GPR09}
S.~Greenberg, A.~Pascoe, and D.~Randall.
\newblock Sampling biased lattice configurations using exponential metrics.
\newblock In {\em Proceedings of the twentieth Annual ACM-SIAM Symposium on
  Discrete Algorithms}, SODA '09, 2009.

\bibitem{GPRjournal}
S.~Greenberg, D.~Randall, and A.P. Streib.
\newblock Sampling biased monotonic surfaces using exponential metrics.
\newblock {\em Arxiv}, 2017.

\bibitem{HW16}
S.~Haddadan and P.~Winkler.
\newblock Mixing of permutations by biased transposition.
\newblock In {\em 34th Symposium on Theoretical Aspects of Computer Science},
  STACS '17, pages 41:1--41:13, 2017.

\bibitem{HH85}
J.~H. Hester and D.~S. Hirschberg.
\newblock Self-organizing linear search.
\newblock {\em Computing Surveys}, 17:295--311, 1985.

\bibitem{Knuth2}
Donald Knuth.
\newblock {\em The Art of Computer Programming}, volume 2: Seminumerical
  Algorithms.
\newblock Addison-Wesley, 1969.

\bibitem{LabbeLacoin}
C.~Labb\'e and H.~Lacoin.
\newblock Cutoff phenomenon for the asymmetric simple exclusion process and the
  biased card shuffling.
\newblock {\em Arxiv:1610.07383v1}, 2016.

\bibitem{LevinPeres}
D.A. Levin and Y.~Peres.
\newblock Mixing of the exclusion process with small bias.
\newblock {\em Journal of Statistical Physics}, 165:1036--1050, 2016.

\bibitem{LRS}
M.~Luby, D.~Randall, and A.J. Sinclair.
\newblock {M}arkov chain algorithms for planar lattice structures.
\newblock {\em SIAM Journal on Computing}, 31:167--192, 2001.

\bibitem{madr}
N.~Madras and D.~Randall.
\newblock Markov chain decomposition for convergence rate analysis.
\newblock {\em Annals of Applied Probability}, pages 581--606, 2002.

\bibitem{mr00}
R.~Martin and D.~Randall.
\newblock Sampling adsorbing staircase walks using a new markov chain
  decomposition method.
\newblock In {\em Proceedings of the 41st IEEE Symposium on Foundations of
  Computer Science}, pages 492--502, 2000.

\bibitem{mr}
R.~Martin and D.~Randall.
\newblock Disjoint decomposition of markov chains and sampling circuits in
  cayley graphs.
\newblock {\em Combinatorics, Probability and Computing}, 15:411--448, 2006.

\bibitem{PR09}
A.~Pascoe and D.~Randall.
\newblock Self-assembly and convergence rates of heterogenous reversible growth
  processes.
\newblock In {\em Foundations of Nanoscience}, 2009.

\bibitem{RT98}
D.~Randall and P.~Tetali.
\newblock Analyzing {G}lauber dynamics by comparison of {M}arkov chains.
\newblock {\em Journal of Mathematical Physics}, 41:1598--1615, 2000.

\bibitem{RM03}
Dana Randall.
\newblock Mixing.
\newblock In {\em Proceedings of the 44th annual IEEE symposium on Foundations
  of Computer Science}, 2003.

\bibitem{sinclair}
Alistair Sinclair.
\newblock {\em Algorithms for random generation and counting}.
\newblock Progress in theoretical computer science. Birkh\"{a}user, 1993.

\bibitem{Spitzer1970}
Frank Spitzer.
\newblock Interaction of {M}arkov processes.
\newblock {\em Advances in Mathematics}, 5:240--290, 1970.

\bibitem{wilson}
D.~Wilson.
\newblock Mixing times of lozenge tiling and card shuffling {M}arkov chains.
\newblock {\em The Annals of Applied Probability}, 1:274--325, 2004.

\end{thebibliography}
\appendix
\section{Appendix}
\subsection{$\mtk$ has the same Stationary Distribution as $\mn$}\label{appendix::sameDist}
In this section we show that the Markov chain $\mtk$ has the same
stationary distribution as $\mn.$  Recall from
Section~\ref{sec::prelim} that the stationary distribution $\pi$ of
$\mn$ is 
$$\pi(\sigma)= \left(\prod_{i<j}p_{\sigma(i), \sigma(j)}\right)Z^{-1},$$ where $Z$
is the normalizing constant $\sum_{\sigma \in \Omega}\prod_{i<j}p_{\sigma(i), \sigma(j)}.$
First, we will show that the transitions of $\mn$ are a subset of the
transitions of $\mtk$ and thus $\mtk$ is also irreducible and connects
the state space $\Omega$.  Consider any nearest neighbor move of
$\mn$, exchanging $\sigma_t(i-1)$ and $\sigma_t(i).$  If
$\C(\sigma_t(i-1)) = \C(\sigma_t(i)),$ then selecting position $i$ and
direction $N$ will result in $\mtk$ making this exchange.  If
$\C(\sigma_t(i-1)) > \C(\sigma_t(i-1))$ then selecting position $i$
and $L$ will result in the move.  Similarly, if $\C(\sigma_t(i-1))
< \C(\sigma_t(i-1))$ then selecting position $i-1$ and $R$ will result
in the move.  

Next, we will show that $\pi$ is the stationary
distribution of $\mtk$ by showing that $\mtk$ satisfies the detailed
balance equations $\pi(\sigma)\pk(\sigma,\tau)
= \pi(\tau)\pk(\tau,\sigma)$ for all $\sigma,\tau\in \Omega.$  Assume
$\tau$ is obtained from $\sigma$ by swapping the elements in positions
$i$ and $j$.  The definition of $\pi$ above implies that 
\begin{equation}\label{eq::ratio}
 \frac{\pi(\sigma)}{\pi(\tau)}
= \frac{p_{\sigma(i),\sigma(j)}}{p_{\sigma_(j),\sigma(i)}}\prod_{m}\frac{p_{\sigma(i),\sigma(m)}p_{\sigma(m),\sigma(j)}}{p_{\sigma(j),\sigma(m)}p_{\sigma(m),\sigma(i)}},
\end{equation}
where the product is over all positions $m$ between $i$ and $j$.
We will consider two cases corresponding to whether $\sigma$ and
$\tau$ differ by a move with direction $N$ or with direction in $\{L,R\}$.  
First, assume $\sigma$ and $\tau$ differ by a $N$ move. This
implies that $\mtk$ selected position $i$, and element $\sigma(j)$ is
the first element to the left of position $i$ in the same class as $\sigma(i)$.  
Therefore, $p_{\sigma(i),\sigma(j)} = p_{\sigma(j),\sigma(i)} = 1/2$ and
$p_{\sigma(i),\sigma(m)} = p_{\sigma(j),\sigma(m)}$ for all $j<m<i.$
Thus, by (\ref{eq::ratio}), $ \frac{\pi(\sigma)}{\pi(\tau)}= 1.$
From the definition of our chain, each move $\pk(\sigma,\tau)
= \pk(\sigma,\tau) = 1/3n.$  Combining these implies that for all
$\sigma, \tau \in \Omega$ that differ by a $N$ move,  $\pi(\sigma)
=\pi(\tau)$ and $\pk(\sigma,\tau) = \pk(\tau,\sigma)$ and thus
detailed balance is satisfied.   

Next, suppose $\sigma$ and $\tau$ differ by the exchange of two
elements in different classes.  Assume without loss of generality that
$\tau$ is obtained from $\sigma$ by making a $L$
move. This implies that they differ by the exchange of two elements
$\sigma(i)$ and $\sigma(j)$ such that $j<i,$
$\C(\sigma(j)) \geq \C(\sigma(i))$ and $\pk(\sigma,\tau) = 1/3n.$  The
only way to move from $\tau$ to $\sigma$ is to select position $j$ and
direction $R$, thus $$\pk(\tau, \sigma)
= \left(\frac{1}{3n}\right) \frac{p_{\tau(i),\tau(j)}}{p_{\tau(j),\tau(i)}}\prod_{j<m<i}\left[\left(\frac{p_{\tau(i),\tau(m)}}{p_{\tau(j),\tau(m)}}\right)\left(\frac{p_{\tau(m),\tau(j)}}{p_{\tau(m),\tau(i)}}\right)\right].$$
Together with (\ref{eq::ratio}), this implies
\[\frac{\pi(\sigma)}{\pi(\tau)} = \frac{p_{\sigma(j),\sigma(i)}}{p_{\sigma(i),\sigma(j)}}\prod_{j<m<i}\frac{ p_{\sigma(j),\sigma(m)} p_{\sigma(m),\sigma(i)}}{p_{\sigma(i),\sigma(m)} p_{\sigma(m),\sigma(j)}}= \frac{\pk(\tau,\sigma)}{\pk(\sigma,\tau)}, \]
thus satisfying detailed balance.
\subsection{Defining the Markov Chains $\m_1, \m_2,\ldots, \m_{k+1}$}\label{appendix::subChains}
We claim that the chain $\mtk$ is a product of $k+1$ independent Markov chains $\{\m_i\}$.  The first $k$ chains $\m_1, \m_2,\ldots, \m_k$ involve moves between particles in the same particle class while the final chain $\m_{k+1}$ allows only moves between different particle classes.  We give the formal definition of each of these chain below.  

\vspace{.1in}
\noindent  {\bf The Unbiased Markov chain $\m_i$ (for $1 \leq i \leq k$) } 

\vspace{.05in}
\noindent {\tt Starting at any permutation $\sigma_0$, iterate the following:} 
\begin{itemize}
\item At time $t,$ choose a position $f$ with $C(\sigma_t(f)) = i$ uniformly at random.  
\item Find the largest $g$ with $1 \leq g < f$ and $C(\sigma_t(g)) =C(\sigma_t(f))= i$.  If such an element exists, then exchange the elements $\sigma_t(f)$ with $\sigma_t(g)$ to obtain $\sigma_{t+1}.$
\item Otherwise, do nothing so that $\sigma_{t+1} = \sigma_{t}.$
\end{itemize}
Next, we define the final chain $\m_{k+1}$ which allows moves between different particle classes. Note that $\m_{k+1}$ includes all moves of $\mtk$ except those with direction $N.$\\

\vspace{.1in}
\noindent  {\bf The Markov chain $\m_{k+1}$} 

\vspace{.05in}
\noindent {\tt Starting at any permutation $\sigma_0$, iterate the following:} 
\begin{itemize}
\item At time $t,$ choose a position $1\leq i \leq n$ and direction $d \in \{L, R\}$ uniformly at random.  
\item If $d = L$, find the largest $j$ with $1 \leq j<i$ and $\C(\sigma_t(j)) \geq \C(\sigma_t(i))$ (if one exists).  If $\C(\sigma_t(j)) > \C(\sigma_t(i)),$ then 
exchange the elements $\sigma_t(i)$ and $\sigma_t(j)$ to obtain $\sigma_{t+1}.$  
\item If $d = R$, find the smallest $j$ with $n\geq j>i$ and $\C(\sigma_t(j)) \geq \C(\sigma_t(i))$ (if one exists).  If $\C(\sigma_t(j)) > \C(\sigma_t(i)),$ then with probability $\lambda_{\sigma_t(j),\sigma_t(i)}\prod_{i<k<j}\left(\lambda_{\sigma_t(j),\sigma_t(k)}\lambda_{\sigma_t(k),\sigma_t(i)}\right),$ exchange the elements $\sigma_t(i)$ and $\sigma_t(j)$ to obtain $\sigma_{t+1}.$  
\item Otherwise, do nothing so that $\sigma_{t+1} = \sigma_t.$
\end{itemize}
\subsection{Bounding the Mixing Time of a Bounded Generalized Exclusion Process}\label{sec::boundedBias}
In this section, we prove Theorem~\ref{boundedBias} from
Section~\ref{sec::exclusion} to analyze bounded generalized exclusion
processes.  We will use the bijection between $\Omega_e$ and staircase
walks given in Section~\ref{sec::exclusion} to view our bounded
generalized exclusion process $\me$ as a ``mountain/valley'' chain over
staircase walks where the probability of adding or removing a square can
vary even for the same square. 

Let $R$ be a rectangular $h \times w$ region in $\mathbb{Z}^2.$
Suppose without loss of generality that $h \leq w.$
Notice $\me$ is equivalent to the following \emph{bounded bias} Markov chain $\ms$ over
staircase walks in $R$: $\ms$ chooses a random diagonal $1\leq i<
h+w$ and a direction (add or remove).  It adds a square above the
staircase walk $S$ along diagonal $i$ with probability 1 if possible, and
removes a square below the staircase walk on diagonal $i$ (if possible) with
probability $p_{\sigma,i}/p_{\sigma',i}$, where $\sigma$ is the
permutation corresponding to $S$ and $\sigma'$ is obtained from
$\sigma$ by swapping $\sigma(i)$ and $\sigma(i+1)$.  Note that
the \emph{bias} $\lambda_{S,(x,y)}:=p_{\sigma',i}/p_{\sigma,i}$ on a
square $(x,y)$ depends on $S$ (and $\sigma$).  Let $\lambda_L$ be the
minimum bias over all squares and all staircase walks, and assume
$\lambda_L>1$ is a constant.
We will prove that the mixing time
$\ts$ of $\ms$ satisfies $$\te(\epsilon) =
O\left(\left(h+w\right)\left(h+\ln w
+\ln \epsilon^{-1}\right)\right).$$

\begin{proof}

Our proof is almost identical to the proof of Theorem~5.2 in the
Greenberg, Streib and Randall~\cite{GPRjournal} paper; we will provide
a high level overview of the proof here.  They prove the following.
First, they show that for staircase walks on rectangular regions with
uniform bias 
$\lambda>1$ where $\lambda$ is a constant, the highest configuration
has constant probability in the stationary distribution.  Next,
they show that for chains with uniform bias $\lambda,$ the mixing time
is $O\left(\left(h+w\right)\left(h+\ln w
+\ln \epsilon^{-1}\right)\right).$  Combining these implies that we
expect the uniform bias chain to hit the highest configuration
in $$O\left(\left(h+w\right)\left(h+\ln w
+\ln \epsilon^{-1}\right)\right)$$ steps.  Moreover,
the \emph{fluctuating bias} chain, where the minimum bias is a
constant $\lambda>1$ will hit the highest configuration \emph{even
faster} than the uniform bias chain with bias $\lambda$.
Specifically they prove the
following. 
\begin{theorem}[\cite{GPRjournal}] Let $R$ be a rectangular $h \times w$ region in $\Z^2$ with fluctuating bias.  Suppose the minimum bias $\lambda_L$ is a constant larger than 1.  Suppose without loss of generality that $h \leq w.$   Then the mixing time of $\mathcal{M}_B$ satisfies $$\tau(\epsilon) = O\left(\left(h+w\right)\left(h+\ln w +\ln \epsilon^{-1}\right)\right).$$
\end{theorem}
Just like with the fluctuating bias Markov chain, the bounded bias Markov chain $\me$ will always lie above the uniform bias chain and so it will also hit the highest configuration in $$O\left(\left(h+w\right)\left(h+\ln w +\ln \epsilon^{-1}\right)\right)$$ steps.  In Greenberg, Streib and Randall~\cite{GPRjournal} Theorem 4.6 they prove that the hitting time to the highest configuration gives an upper bound on the coupling time.  We can apply this theorem directly and thus the mixing time $\ts$ of the bounded bias chain with $\lambda_L$ a constant larger than one satisfies $$\ts(\epsilon) = O\left(\left(h+w\right)\left(h+\ln w +\ln \epsilon^{-1}\right)\right).$$\\
\end{proof}

\subsection{Bounding the Mixing Time of $\mn$ using the Bound on $\mtk$}\label{appendix::comparison}
Recall from Section~\ref{sec::kcomparison} that we will use the comparison theorem (Theorem~\ref{comparison}) to bound the mixing time of $\mn$ using the bound on the mixing time of $\mtk.$  In order to apply the comparison theorem, for each edge $(x,y)$ in $\mtk$ we will need to define a path $\gamma_{xy}$ using edges in $\mn$.  In this section we complete the proof of Theorem~\ref{mnMixing} by formally describing the paths $\gamma_{xy}$ for each edge $(x,y)$.  If $(x,y)$ is a nearest-neighbor swap, then it is also a transition in $\mn$ and we will simply use this single edge as the path (i.e. $\gamma_{xy} = \{x,y\}$).  Thus, we can focus on edges that swap two elements that are not adjacent.  In each step of $\mtk$ we choose a position and then a direction.  We will define different paths based on the direction.  For each type of path we will need to bound the length of each path, and for each transition $(z,w)$ of $\mn$, we will need to bound $|\Gamma(z,w)|$, the number of paths $\gamma_{x,y}$ that pass through $(z,w)$.  

First consider the $R$ direction.  Without loss of generality, assume $x$ and $y$ differ by an $(a,b)$ transposition where $x(i) = a$,  $x(j)=b$, $i<j$ and $\C(b) > \C(a).$  We will define the path $\gamma_{xy} = \{x=x_0,x_1,x_2,\cdots x_s = y\}$ of transitions $(x_i,x_{i+1})$ of $\mn$  as follows.  Our path will first take element $a$ and swap it to the right until it is at position $j-1.$  Next we will exchange $a$ and $b$ and finally move $b$ to the left until it is at position $i$.  Since $j-i < n,$ the length of this path is at most $2n.$  Recall that by the definition of the direction $R$ moves, for every element $c$ between $i$ and $j$, $\C(c)<\C(a) < \C(b)$ and thus by moving $x(i)$ to the right first,we ensure that at each step  the weight of the configuration remains at least the weight of the original configuration.  Given a particular edge in the path, if we know the direction of the move was $R$ and particle $a$'s original location, then we know exactly which configurations we are moving between.  Hence, each edge can be associated with at most $n+1$ different ``R" paths.  Note that the $L$ direction is very similar and so we defer the proof to the full version of the paper.

Next, suppose the direction selected is $N.$   In Section~\ref{sec::kcomparison} we gave a high-level description of the paths in this case.  Assume $a$ is in position $i$ and $b$ is in position $j>i$.  The complete paths for direction $N$ are formally described in Figure~\ref{canonicalPaths}.  
To complete the proof, it remains to bound the length $|\gamma_{x,y}|$ of any ``N" path, show that the path does not go through any permutations of small weight, and finally that not too many paths go through any particular edge of $\mn$.
\begin{figure}[t]
\begin{center}
\begin{minipage}{0.8\textwidth}
\underline{The Canonical Path for a Direction $N$ Edge in $\mtk$}\\

{\bf Phase I }(move element $b$ to the left):\\
$i \gets 0$ (stores the current step in the path $x_0,\ldots, x_s$)\\
$j \gets$ the location of element $b$ in $x=x_0$\\
REPEAT\\
\myInd IF $x(j)=a$ ($b$ is in the original location of $a$) \\
\myInd\myInd GOTO Phase II \\
\myInd IF $\C(x_{i}(j-1)) >= \C(b)$ \\
\myInd\myInd exchange the elements $x_i(j)$ and $x_i(j-1)$ to obtain $x_{i+1}$\\
\myInd\myInd $i \gets i+1, j \gets j-1$\\
\myInd ELSE\\
\myInd\myInd  $l \gets \max\{l < j: \C(x_i(l)) >= \C(b)\}$  \\
\myInd\myInd FOR $m \in \{l,\ldots j-2\}$ \\
\myInd\myInd \myInd exchange the elements $x_i(m)$ and $x_i(m+1)$ to obtain $x_{i+1}$ \\
\myInd\myInd \myInd $i \gets i+1$\\
\myInd\myInd FOR $m \in \{j-1,\ldots l\}$ \\
\myInd\myInd \myInd exchange elements $b$ and $x_i(m)$ to obtain $x_{i+1}$ \\
\myInd\myInd \myInd $i \gets i+1, j \gets j-1$\\

{\bf Phase II} (move element $a$ to the right):\\
$j \gets$ the location of element $a$ in $x_i$. \\
REPEAT\\
\myInd IF $x(j) = b$ ($a$ is in the original location $b$)\\
\myInd\myInd STOP \\
\myInd IF $\C(x_{i}(j+1)) > \C(a)$  \\
\myInd\myInd exchange the elements $x_i(j)$ and $x_i(j+1)$ to obtain $x_{i+1}$\\
\myInd\myInd $i \gets i+1, j \gets j+1$\\
\myInd ELSE\\
\myInd\myInd $l \gets \min\{l > j: \C(x_i(l)) > \C(a) \}$  \\
\myInd\myInd FOR $m \in \{k+1,\ldots l\}$ \\
\myInd\myInd \myInd exchange elements $a$ and $x_i(m)$ to obtain $x_{i+1}$ \\
\myInd\myInd \myInd $i \gets i+1, j \gets j+1$\\
\myInd\myInd FOR $m \in \{l-2,\ldots j\}$ \\
\myInd\myInd \myInd exchange the elements $x_i(m)$ and $x_i(m+1)$ to obtain $x_{i+1}$ \\
\myInd\myInd \myInd $i \gets i+1$
\end{minipage}
\end{center}
\caption{The path $\gamma_{xy}=\{x=x_0,x_1,\ldots, x_s=y\}$ exchanges elements $a$ and $b$ with $\C(a) = \C(b).$}
\label{canonicalPaths}
\end{figure}

First we will bound the length $|\gamma_{x,y}|$ of any path of this type. 
In our path $\gamma_{x,y},$ elements $a$ and $b$ are each involved in exactly $j-i$ nearest neighbor transpositions, giving a total of $2j-2i-1$ transpositions (because one involves both $a$ and $b$).  Additionally we may have transpositions involving neither $a$ or $b$.  Due to the construction of the paths, these transpositions are all between an element in a particle class greater than $\C(a)$ and an element in a particle class smaller than $\C(a).$  Additionally, each element in a particle class greater than $\C(a)$ is involved in at most 2 of these transpositions (one while moving $b$ to the left and one while moving $a$ to the right).  Thus there are at most $2(j-i-1)$ of these transpositions.   Combining these observations with $j-i-1 \leq n-2$ ($n$ total elements thus at most $n-2$ elements between $a$ and $b$) shows that the total number of transpositions $|\gamma_{x,y}| \leq 4(j-i-1)+1 \leq 4n.$

Next, we'll show that the weight of any configuration  $x_i$ on the path $\gamma_{xy} = \{x=x_0,x_1,x_2,\cdots x_s = y\}$ satisfies $\pi(x_i) \geq\min (\pi(x), \pi(y)).$  First, let's consider the case where $(x,y)$ corresponds to a $N$ move and we will again assume we are exchanging two elements $a$ and $b$ and consider the steps in Phase I of the path (Phase II is almost identical).  The only steps in Phase I that do not increase the weight of the configuration are those that exchange $b$ with an element whose particle class is smaller than $\C(b).$  Consider any maximal contiguous block of $k$ such elements $e_1, \ldots e_k$.  Before we move element $b$ across these elements we first move a different element $d$ with $\C(d) > \C(b)$ to the right across these elements.  When element $b$ is then moved across these $k$ elements, the elements in the middle return to their original position and the overall weight is increased by a factor of $\prod_{j=1}^k \frac{p_{b,e_i}p_{e_i,d}}{p_{d,e_i}p_{e_i,b}}\geq 1,$ since the probabilities are weakly monotonic (property 2)\footnote{If the probabilities $\p$ satisfy the weak monotonicity property 3 instead then we would need to modify the path to first move element $a$ to the right and then move element $b$ to the left.  Additionally, as mentioned previously we would also need to update $\mtk$ to allow swaps between elements in different particle classes across elements whose particle class is larger (instead of smaller).  The paths for direction $L$ and $R$ would need to be updated accordingly.}. 

Next, we need to bound the number of canonical paths of type ``N" that use an edge $(z,w)$ of $\mn$.  We'll show that for any edge $(z,w)$ with $\pn(z,w) > 0$, $|\Gamma(z,w)| \leq 6n^3.$ Here will use an information theoretic argument.    
Let $L$ be the set of elements between $a$ and $b$ in a larger particle class than $\C(a) = \C(b)$ and similarly let $S$ be the set in a smaller particle class than $\C(a).$  Notice that throughout the path the relative order of the elements in $L$ and in $S$ remains the same.  If we consider the placement of elements in $S$ versus elements in $L$ (i.e. replace elements in $L$ with 0's and elements in $S$ with 1's) then this also remains almost the same with at most one element out of place at any point in the path.  Given these observations, it is fairly straightforward to see that given any edge $(z,w),$ as long as we have the following information we can recover the two configurations $x,y$ completely and thus this gives a bound on the number of pairs of configurations that use a particular edge. 
\begin{enumerate}
\item The direction of the move (3)
\item The original location of element $a$ or $b$ (whichever is not in it's original, or final position). ($n$)
\item Whether the move involves element $a$ (or $b$).(2)
\item If the move does involve element $a$ (or $b$) then which of the 2 elements being exchanged is $a$ or $b.$ (2)
\item If the move does not involve element $a$ (or $b$) then the original and final location of the element in $L$ that is being moved. ($n^2$)
\end{enumerate}
If the direction is not $N$ then as described previously we actually need less information to recover the original $\mtk$ edge $(x,y).$
Thus the number of possible paths using a particular edge is at most~$6n^3.$

Finally, we can bound the quantity $A$ from the comparison theorem and prove Theorem~\ref{mnMixing}.  Regardless of the direction ($L$,$R$ or $N$), the upper bound on the length of the canonical paths ($|\gamma_{x,y}| \leq 4n$) gives the following.
\begin{eqnarray*}
A &=& \max_{(z,w) \in E(\pn)} \left \{\frac{1}{\pi(z)\pn(z,w)}\sum_{\Gamma(z,w)}|\gamma_{xy}|\pi(x)\pk(x,y) \right \} \\
&\leq& \max_{(z,w) \in E(\pn)} \left \{4n\sum_{\Gamma(z,w)}\left(\frac{\pi(x)}{\pi(z)}\right)\left(\frac{\pk(x,y)}{\pn(z,w)}\right) \right \}.
\end{eqnarray*}

Let us assume $x$ and $y$ differ by a single transposition of two elements $a = x(i)$ and $b=x(j)$ with $i < j$.  We will consider three cases depending on the direction of the $(x,y)$ move and in each case show that  $\left(\frac{\pi(x)}{\pi(z)}\right)\left(\frac{\pk(x,y)}{\pn(z,w)}\right)\leq  1/3\pmin,$  where $\pmin$ is the minimum $p_{i,j}$ for $i>j.$  First, if $(x,y)$ is a direction $N$ move then this implies that $\pi(y) = \pi(x).$  Since $\pi(z) \geq \min(\pi(x), \pi(y)) = \pi(x),$ we have $\pi(x)/\pi(z) \leq 1.$  Additionally, $\pk(x,y) =1/3n$ and  thus $(\pk(x,y)/\pn(z,w)) \leq 1/3\pmin.$
  Combining these gives an upper bound $\left(\frac{\pi(x)}{\pi(z)}\right)\left(\frac{\pk(x,y)}{\pn(z,w)}\right)\leq  1/3\pmin,$ as desired.  Similarly if the direction is $L$ this implies that $\pi(y) > \pi(x).$   Again since $\pi(z) \geq \min(\pi(x), \pi(y)) = \pi(x),$ we have $\pi(x)/\pi(z) \leq 1,$  $\pk(x,y) =1/3n$ and thus $(\pk(x,y)/\pn(z,w)) \leq 1/3\pmin$ and again  $\left(\frac{\pi(x)}{\pi(z)}\right)\left(\frac{\pk(x,y)}{\pn(z,w)}\right)\leq  1/3\pmin.$
If the direction is $R$ then $\pi(y) < \pi(x)$ and $\pi(z) \geq \min (\pi(x), \pi(y))$ implies $\pi(z) \geq \pi(y)$  and thus  $\pi(x)/\pi(z) \leq \pi(x)/\pi(y) = \frac{p_{a,b}}{p_{b,a}}\prod_{i<k<j}\frac{p_{a,x(k)}p_{x(k),b}}{p_{b,x(k)}p_{x(k),a}}.$  From the definition of $\mtk$ we have that $\pk(x,y) = (1/3n)\frac{p_{b,a}}{p_{a.b}}\prod_{i<k<j}\frac{p_{b,x(k)}p_{x(k),a}}{p_{a,x(k)}p_{x(k),b}}$ and thus again in the $R$ case, $\left(\frac{\pi(x)}{\pi(z)}\right)\left(\frac{\pk(x,y)}{\pn(z,w)}\right)\leq  1/3\pmin.$ Combining these with the bound on $|\Gamma(z,w)| $ gives $$A \leq \max_{(z,w) \in E(\pn)}  4n \frac{|\Gamma(z,w)|}{\pn(z,w)} \leq 72n^4\pmin.$$

Using this bound on $A$, our bound on the minimum weight configuration
$\pi_*$ from Section~\ref{sec::kparticle} and appealing to the
comparison theorem (Theorem~\ref{comparison}) and our bound on the
mixing time of $\mtk$ (Theorem~\ref{mkMixing}) gives the following,
\begin{align*}
\tn(\epsilon) &= O\left( (n^2\log \epsilon^{-1})
(n^4 \pmin)(n^{2k} \log(k\epsilon^{-1})\right)/ \log(1/2\epsilon)\\
&= O\left( n^{2k+6}\log(k\epsilon^{-1})\right).
\end{align*}

\noindent If we let $\epsilon = 1/4$ then we have that $\tn = O(n^{2k+6}\log k).$

\subsection{Bounding the Mixing Time of $\mpp$ using the Bound on $\m_{k+1}$}\label{appendix::particleprocess}
In this section we provide the details to complete the proof of Corollary~\ref{mpMixing} which bounds the mixing time of the particle process chain $\mpp$ where particles in a single class are indistinguishable.  Note that since particles are indistinguishable we will bound the mixing time of $\mpp$ again using the comparison theorem (Theorem~\ref{comparison}) and the bound on the mixing time of $\m_{k+1}$ (Lemma~\ref{pprocess}).  Recall that the moves of $\m_{k+1}$ are the moves of $\mtk$ with direction $L$ and direction $R.$  Thus, we will use the exact same paths define above (Section~\ref{appendix::comparison}) for the $L$ and $R$ directions.  As explained previously, using these canonical paths a single edge can be used in at most $n+1$ different $R$ paths and $n+1$ different $L$ paths.  The bound on the maximum length of a path remains at most $4n$.  Combining these gives a bound on $A$ of $O(n^2)$.  Using this bound on $A$, our bound on the minimum weight configuration $\pi_*$ from Section~\ref{particleToClass} and appealing to the comparison theorem (Theorem~\ref{comparison}) and our above bound on the mixing time of $\m_{k+1}$ (Lemma~\ref{pprocess}) gives 
$\tp(\epsilon)=  O\left(n^{2k+4}\ln(\epsilon^{-1})\right).$

\subsection{Tree Representation}\label{appendix::tree}
In this section, we fill in details left out of Section~\ref{sec::tree}.  First, we define the bijection between tree representations and permutations.  To obtain the tree representation from the permutation, perform the following.

\begin{itemize}
  \item For each non-leaf node $v$ do the following:
    \begin{itemize}
      \item List each leaf descendant $x$ of $v$ in the order we encounter them
        in the permutation $\sigma$ (all leaf nodes in $v$'s subtree).
      \item For each listed element $x$, write the label of node $a$ where $a$ is the child of $v$ such that $x$ is contained in the subtree rooted at $a.$ 
    \end{itemize}
\end{itemize}

\noindent We see that any $\sigma$ will lead to an assignment of strings as described above at each
non-leaf node $v$.   Given any tree representation, we can
recursively reconstruct the permutation $\sigma$ as follows:

\begin{itemize}
  \item For each leaf node $i$, let its permutation string be the string $``i"$.
  \item For any node $v$ with tree string $s$,
    \begin{itemize}
      \item Determine the permutation strings of each of its children. Call these $s_1, \ldots, s_{deg(v)}$.
      \item Create $v$'s permutation string by interleaving the elements of $s_1,\ldots, s_{deg(v)}$ in the relative order specified by $s.$   Specifically, working from left to right, replace each element $i$ in $s$ with the next element
        from $s_i.$  The elements from $s_i$  should remain in the same relative order in $v$'s permutation string as they are originally in $s_i.$
    \end{itemize}
\end{itemize}

\end{document}